\documentclass[12pt]{article}
\usepackage{epsfig,amsmath,amssymb,mathrsfs}

  {\end{list}}%

\makeatletter
\renewcommand{\section}{\setcounter{equation}{0}\@startsection
 {section}%
 {1}%
 {0pt}%
 {-1\baselineskip}%
 {0.4\baselineskip}%
 {\bfseries\large}}%
\renewcommand{\subsection}{\@startsection
 {subsection}%
 {2}%
 {0pt}%
 {-0.75\baselineskip}%
 {0.2\baselineskip}%
 {\bfseries}}%
\renewcommand{\subsubsection}{\@startsection
 {subsubsection}%
 {3}%
 {0pt}%
 {-0.5\baselineskip}%
 {0.1\baselineskip}%
 {\sc}}%
\makeatother

\DeclareMathAlphabet{\mathpzc}{OT1}{pzc}{m}{it}

\def\g5{\gamma_{5}}

\def\idpi{\frac{d^4\!p_i}{(2\pi)^4}}
\def\idpone{\frac{d^4\!p_1}{(2\pi)^4}}
\def\idptwo{\frac{d^4\!p_2}{(2\pi)^4}}
\def\idpthree{\frac{d^4\!p_3}{(2\pi)^4}}

\def\idx{\int\!\! d^4\!x}

\newcommand{\bea}{\begin{eqnarray}}
\newcommand{\eea}{\end{eqnarray}}
\newcommand{\beann}{\begin{eqnarray*}}
\newcommand{\eeann}{\end{eqnarray*}}
\newcommand{\ba}{\begin{array}}
\newcommand{\ea}{\end{array}}

\newcommand{\Tr}{\mathbf{Tr}}

 \def\g {\gamma}

\begin{document}
 \begin{titlepage}
\rightline{FTI/UCM 153-2013}
\vglue 33pt

\begin{center}

{\Large \bf SO(10) GUTs with large tensor representations on Noncommutative Space-time}\\
\vskip 1.0true cm
{\rm C. P. Mart\'{\i}n}\footnote{E-mail: carmelop@fis.ucm.es}
\vskip 0.1 true cm
{\it Departamento de F\'{\i}sica Te\'orica I,
Facultad de Ciencias F\'{\i}sicas\\
Universidad Complutense de Madrid,
 28040 Madrid, Spain}\\

\end{center}

{\leftskip=50pt \rightskip=50pt \noindent
We construct a noncommutative version of  a general renormalizable SO(10) GUT with Higgses in the 210, $\overline{126},  45,  10$ and 120
irreps of SO(10) and a Peccei-Quinn symmetry. Thus, we formulate the  noncommutative counterpart of a non-supersymmetric SO(10) GUT which has recently been shown
to be consistent with all the physics below $M_{GUT}$. The simplicity of our construction --the simplicity of the Yukawa terms, in particular-- stems from the fact that the Higgses of our
GUT can be viewed as elements of the Clifford algebra $\mathbb{C}\rm{l}_{10}(\mathbb{C})$; elements on  which the SO(10) gauge transformations act by conjugation. The noncommutative GUT we
build contains tree-level interactions among different Higgs species that are absent in their ordinary counterpart as they are forbidden by SO(10)  and Lorentz
invariance. The existence of these interactions   helps to clearly distinguish noncommutative Minkowski space-time from ordinary Minkowski space-time.
\par}

\vspace{9pt}
\noindent{\em PACS:} 11.10.Nx; 12.10.-g; \\
{\em Keywords:}  Noncommutative gauge theories, GUTs.
\vfill
\end{titlepage}

\section{Introduction}

It has recently been shown in Ref.~\cite{Altarelli:2013aqa} that a non-supersymmetric SO(10) GUT with Higgses in the 210, $\overline{126}$, 45 and 10 irreps of SO(10) --and with an intermediate breaking to the Patti-Salam group by the 210--  is compatible with all the experimental data currently available, if the naturalness paradigm is put aside. A salient feature of this GUT is that the amount of dark matter that has been observed  is accounted for by the existence of an axion which results from the spontaneously broken Peccei-Quinn symmetry of the theory. Under this global U(1) symmetry some Higgses are charged, others  are not.

It is almost 15 years~\cite{Seiberg:1999vs} since it became well-established that ordinary Minkowski space-time might have to be replaced with its noncommutative counterpart as one probes shorter distances. Hence, it is interesting to see whether there can be constructed on noncommutative space-time a field theory which can be considered  to be a noncommutative version of the phenomenologically relevant  SO(10) GUT of Ref.~\cite{Altarelli:2013aqa}.
The purpose of this paper is to show that, indeed, a noncommutative counterpart of the ordinary GUT in Ref.~\cite{Altarelli:2013aqa} can be formulated. We shall actually  enlarge, for the sake of generality, the Higgs content of that GUT with a Higgs in the 120 irrep of SO(10), for the latter  naturally occurs in the most general ordinary SO(10) Yukawa term  for fermions in the 16. Ordinary SO(10) GUTs with Higgses only in the 210, $\overline{126}$, 45, 10 an 120 are very suitable for their generalization to GUTs on noncommutative space-time, for the Higgses they  involve can naturally be understood as  elements of the Clifford algebra $\mathbb{C}\rm{l}_{10}(\mathbb{C})$ and SO(10) acts on these elements by conjugation. This feature of the Higgses --which is very appealing from the noncommutative geometry standpoint~\cite{GraciaBondia:2001tr}-- is lost if one considers Higgses in the 16 or 54 irreps of SO(10), another popular Higgs irreps in SO(10) model building.

The formulation of the noncommutative counterpart of the SO(10) GUT of Ref.~\cite{Altarelli:2013aqa} will be carried out with the help of the enveloping-algebra formalism. This formalism was put forward in Refs.~\cite{Madore:2000en, Jurco:2000ja,
  Jurco:2001rq}. The enveloping-algebra framework  was employed afterwards to build the noncommutative Standard Model~\cite{Calmet:2001na}, a noncommutative deformation of the ordinary Standard Model with no new degrees of freedom --see Refs.~\cite{Chaichian:2001py, Khoze:2004zc} for
alternative noncommutative extensions of the ordinary Standard Model. The formulation of the gauge and fermionic sectors of  noncommutative GUTs with SU(5) and SO(10) as gauge groups was tackled, within the enveloping-algebra framework,  in Ref.~\cite{Aschieri:2002mc}. The nontrivial issue of constructing noncommutative Yukawa terms with the help of the enveloping-algebra formalism was addressed in Ref.~\cite{Martin:2010ng}. Outside the enveloping-algebra framework, the formulation of noncommutative gauge theories for SO(N) groups was  discussed in Ref.~\cite{Bonora:2000td}.

In the enveloping-algebra framework, the noncommutative gauge fields are elements of the universal enveloping algebra of the Lie algebra of the ordinary gauge group and the Seiberg-Witten map defines those noncommutative fields in terms of the corresponding ordinary fields.  When the Seiberg-Witten map is defined  as a formal power series in the noncommutativity matrix parameter $\omega^{\mu\nu}$, the action of the noncommutative theory is  a formal power series in $\omega^{\mu\nu}$ with coefficients that are integrated polynomials in the ordinary fields and their derivatives.
Quite a few theoretical properties  --e.g., renormalizability~\cite{Buric:2005xe, Buric:2006wm, Buric:2007ix, Martin:2007wv, Martin:2009sg, Tamarit:2009iy, Martin:2009vg}, gauge anomalies~\cite{Martin:2002nr, Brandt:2003fx}, existence of noncommutative deformations of ordinary instantons and monopoles~\cite{Martin:2005vr, Martin:2006px, Stern:2008wi}-- of the noncommutative gauge theories so defined have been analyzed by considering the first few terms of the corresponding  $\omega^{\mu\nu}\!$-expanded actions. Some phenomenological properties of the noncommutative gauge theories at hand have been  studied in Refs.~\cite{Melic:2005su, Alboteanu:2006hh, Buric:2007qx, Tamarit:2008vy,Trampetic:2009vy, Haghighat:2010up,  YaserAyazi:2012ni, Aghababaei:2013dia}.

The UV/IR mixing effects~\cite{Minwalla:1999px} that are a feature of  the $\omega^{\mu\nu}\!$-unexpanded U(N) noncommutative field theories cannot be exhibited in a noncommutative gauge theory built with the help of
Seiberg-Witten map, when this map is defined as a series expansion in $\omega^{\mu\nu}\!$. To uncover such UV/IR effects in this $\omega^{\mu\nu}\!$-expanded theories  some kind of re-summation of an infinite number of terms that are powers of  $\omega^{\mu\nu}$ must be  worked out: a daunting task. Fortunately, for the enveloping-algebra formalism to work~\cite{Jurco:2001rq} it is not a must  that the Seiberg-Witten be given by a formal series expansion in $\omega^{\mu\nu}$. Indeed, the enveloping-algebra framework works equally well if the Seiberg-Witten map is defined by expanding in the number of ordinary fields, thus leaving its dependence on $\omega^{\mu\nu}$ exact. Hence, to study noncommutative UV/IR effects in theories defined within the enveloping-algebra formalism one should use this $\omega^{\mu\nu}\!$-exact
Seiberg-Witten map. This was done for the first time in Ref.~\cite{Schupp:2008fs} were it was shown, in the $U(1)$ case with fermions in the adjoint, that if the $\omega^{\mu\nu}$ dependence of the Seiberg-Witten is handled exactly, then, there is an UV/IR mixing phenomenon in the
noncommutative theory defined within the enveloping-algebra formalism. The analysis of the UV/IR mixing effects was later
extended~\cite{Raasakka:2010ev} to fermions in the fundamental representation coupled to $U(1)$ gauge fields. Very recently the complete one-loop photon and neutrino propagators have been worked out and its full UV/IR mixing structure unveiled --see Ref.~\cite{Horvat:2013rga} . The UV/IR mixing in the
one-loop propagator of adjoint fermions coupled to U(1) fields and its very relevant implications on neutrino physics have been studied in
Refs.~\cite{Horvat:2011iv, Horvat:2011bs, Horvat:2011qn, Horvat:2011qg} --see
Ref.~\cite{Trampetic:2013ey} for a review.  Finally, let us stress that the cohomological techniques developed in
Refs.~\cite{Barnich:2001mc, Barnich:2003wq} --see also~\cite{Ulker:2007fm}--
are extremely useful~\cite{Martin:2012aw} when computing  the ($\omega^{\mu\nu}\!$-exact)
expansion of the Seiberg-Witten map in the number of ordinary fields.

As we said at the beginning of this introduction, the purpose of this paper is to show that there is indeed a noncommutative
counterpart of the SO(10) GUT of Ref.~\cite{Altarelli:2013aqa}. We shall give the complete action of the noncommutative SO(10) GUT: its Yukawa and Higgs parts, in particular. The action will be expressed in terms of noncommutative fields whose noncommutative gauge transformations are the natural  generalization  of the corresponding ordinary gauge transformations. That this strategy works for the Yukawa and Higgs terms of our SO(10) GUT is a consequence of the fact that the ordinary Higgses of our theory can be viewed as elements of $\mathbb{C}\rm{l}_{10}(\mathbb{C})$ and that the gauge transformations act on these objects by conjugation. An important by-product  of this  Clifford algebra construction is that the noncommutative Higgs action naturally contains terms which can not occur when space-time is commutative and there is Lorentz invariance. These distinct noncommutative terms give rise to tree-level interactions among different species of Higgses and gauge fields, and if experimentally detected will send a clear signal that space-time is noncommutative at short enough distances.

The layout of this paper is as follows. In Section 2, we display the field content and gauge transformations of the ordinary SO(10) GUT whose noncommutative version we shall construct afterwards.
The noncommutative fields of our SO(10) GUT, along with the Seiberg-Witten map equations that define them in terms of their ordinary counterparts,  are given in Section 3. The action of the Peccei-Quinn symmetry on the
noncommutative fields is discussed in Section 4. Section 5 is devoted to the construction of the action of our noncommutative SO(10) GUT. Some future research directions  are given Section 6. We include an appendix where
the $\omega$-exact Seiberg-Witten map for the Higgs fields viewed as elements of $\mathbb{C}\rm{l}_{10}(\mathbb{C})$ is given up to order 2 in the number of gauge fields.

\section{The field content of the ordinary SO(10) GUT and its gauge invariance}

Let us list   the matter field content of our ordinary SO(10) GUT, a particular instance of which is the
SO(10) GUT of Ref.~\cite{Altarelli:2013aqa}.
   First, three --one for each
family in the Standard Model--, left-handed fermionic fields
$\psi^{(16)\,f}_{\alpha},\,f=1,2,3$, transforming under the 16 irrep of SO(10) and the (1/2,0) representation of the Lorentz group. Each
$\psi^{(16)\,f}_{\alpha}$ contains the  fermionic fields of a family of the Standard Model plus the degrees of freedom corresponding a right-handed neutrino. Secondly, five Higgs fields, namely, $\varphi^{(210)}_{i_1i_2i_3i_4}, \varphi^{(10)}_{i_1}, \varphi^{(45)}_{i_1i_2}, \varphi^{(\overline{126})}_{i_1i_2i_3i_4i_5}$,
and $\varphi^{(120)}_{i_1i_2i_3}$, carrying, respectively,  the 210, the
10, the 45, the $\overline{126}$ and the 120  irreps of SO(10). The indices $i_1,i_2,....$ run
from 1 to 10,  and  $\varphi^{(210)}_{i_1i_2i_3i_4}$,  $\varphi^{(\overline{126})}_{i_1i_2i_3i_4i_5}$
and $\varphi^{(120)}_{i_1i_2i_3}$ are totally antisymmetric,  with regard to its  $i_1, i_2,...$ indices, SO(10) tensors. Further, $\varphi^{(\overline{126})}_{i_1i_2i_3i_4i_5}$
satisfies the following  duality equation:
\begin{equation}
\varphi^{(\overline{126})}_{i_1i_2i_3i_4i_5}=+\frac{i}{5!}\,\varepsilon_{i_1i_2i_3i_4i_5i_6i_7i_8i_9i_{10}}\,\varphi^{(\overline{126})}_{i_6i_7i_8i_9i_{10}}.
\label{selfdual}
\end{equation}

The symbol $\varphi^{(H)}_I$, $I=1...\,dim\; H$, $H= 210, 10, 45, \overline{126}$ and $120$, will stand for the independent components of   $\varphi^{(210)}_{i_1i_2i_3i_4}, \varphi^{(10)}_{i_1}, \varphi^{(45)}_{i_1i_2}, \varphi^{(126)}_{i_1i_2i_3i_4i_5}, \varphi^{(\overline{126})}_{i_1i_2i_3i_4i_5}$,
and $\varphi^{(120)}_{i_1i_2i_3}$, respectively. $dim\;H$ is the dimension of the representation $H$.

The gauge field content of our GUT is furnished by the 45 gauge fields $a_{\mu}^{i j}$, with $a_{\mu}^{i j}=-a_{\mu}^{j i}$ and $i,j=1...10$, which constitute the 45 irrep of SO(10).

Let $\Gamma^{i}$ denote the hermitian Dirac matrices in 10 Euclidean dimensions. These
matrices generate the Clifford algebra $\mathbb{C}\rm{l}_{10}(\mathbb{C})$.
We shall see later that  noncommutative counterparts of the Yukawa terms and some Higgs potential terms of our ordinary SO(10) GUT can be formulated
very neatly by using the $\mathbb{C}\rm{l}_{10}(\mathbb{C})$ Clifford algebra valued Higgs fields:
\begin{equation}
\begin{array}{l}
{\phi^{(210)}=\Gamma^{i_1}\Gamma^{i_2}\Gamma^{i_3}\Gamma^{i_4}\varphi^{(210)}_{i_1i_2i_3i_4},\;
\phi^{(10)}=\Gamma^{i_1}\varphi^{(10)}_{i_1},\;
\phi^{(45)}=i\Gamma^{i_1}\Gamma^{i_2}\varphi^{(45)}_{i_1i_2},}\\[4pt]
{\phi^{(\overline{126})}=\Gamma^{i_1}\Gamma^{i_2}\Gamma^{i_3}\Gamma^{i_4}\Gamma^{i_5}\varphi^{(\overline{126})}_{i_1i_2i_3i_4i_5},\;
\phi^{(120)}=i\Gamma^{i_1}\Gamma^{i_2}\Gamma^{i_3}\varphi^{(120)}_{i_1i_2i_3},}
\end{array}
\label{Clifhiggs}
\end{equation}
rather than the SO(10) tensor fields
$\varphi^{(210)}_{i_1i_2i_3i_4}$, $\varphi^{(10)}_{i_1}$, $\varphi^{(45)}_{i_1i_2}$,
$\varphi^{(\overline{126})}_{i_1i_2i_3i_4i_5}$ and $\varphi^{(120)}_{i_1i_2i_3i_4}$,
which give rise to the former.

From now on, the symbol $a_\mu$ will stand for the following gauge field taking values in the
in the Lie algebra of SO(10) in the $16\bigoplus\overline{16}$ representation:
\begin{equation}
a_\mu=\frac{1}{2}\Sigma^{ij}a^{ij}_{\mu},\quad\Sigma^{ij}=\frac{1}{4i}\,[\Gamma^i,\Gamma^j],\quad
  i,j=1...10.
\label{ordsvec}
\end{equation}
The real fields $a^{ij}_{\mu}$ carry the 45 irrep of SO(10) and has been introduced above. Notice that $a_\mu$ is an element of the Clifford algebra $\mathbb{C}\rm{l}_{10}(\mathbb{C})$. Besides, $a_\mu=(a_\mu)^{\dagger}$.

Let $M^{(H)}_{ij}$, $i<j,\quad i,j=1...10$,  be the hermitian generators of SO(10) in the representation carried by $\varphi^{(H)}_I$, $I=1...\,dim\; H$. Then, the matrix gauge field $a^{(H)}_{\mu}$ is given by
\begin{equation}
a^{(H)}_{\mu}=\frac{1}{2}a^{ij}_{\mu}\,M^{(H)}_{ij}, \quad (a^{(H)}_{\mu})^{\dagger}=a^{(H)}_\mu.
\label{hafield}
\end{equation}

Let us now introduce the BRS transformations that constitute the gauge symmetry of our GUT. Let $s$ denote the
BRS operator, $c=\frac{1}{2}\Sigma^{ij}c^{ij}$ the ghost field associated to $a_{\mu}$ and $c^{(H)}=\frac{1}{2}M^{(H)}_{ij}c^{ij}$ the ghost field associated to $a^{(H)}_{\mu}$; then, we have the following  BRS transformations
\begin{equation}
\begin{array}{l}
{sc=-icc,\quad sa_{\mu}=D_{\mu}c=\partial_{\mu}c+i[a_{\mu},c],}\\[4pt]
{sc^{(H)}=-ic^{(H)}c^{(H)},\quad sa^{(H)}_{\mu}=D_{\mu}c^{(H)}=\partial_{\mu}c^{(H)}+i[a^{(H)}_{\mu},c^{(H)}],}\\[4pt]
{s\psi^{(16)\,f}_\alpha=-ic\psi^{(16)\,f}_\alpha,\quad s\phi^{(H)}=-i[c,\phi^{(H)}],\quad s\varphi^{(H)}=-i c^{(H)}\varphi^{(H)}. }
\label{ordBRStrans}
\end{array}
\end{equation}
Using the condition $(c^{ij})^{*}=c^{ij}$, one concludes that
\begin{equation*}
s(\psi^{(16)\, f}_\alpha)^{\dagger}=i(\psi^{(16)\, f}_\alpha)^{\dagger}c,\quad s(\phi^{(H)})^{\dagger}=-i[c,(\phi^{(H)})^{\dagger}],\quad s(\varphi^{(H)})^{\dagger}=i(\varphi^{(H)})^{\dagger}c^{(H)}.
\end{equation*}
In the previous equations, and in the sequel, $\psi^{(16)\, f}_\alpha$ is viewed as the projection of a 32 Dirac spinor onto the 16 dimensional Weyl spinor subspace that carries 16 irrep of SO(10). This projection is carried out by the operator $P_{+}=1/2(1+\Gamma_{11})$,  $\Gamma_{11}=i^5\Gamma_1\Gamma_2...\Gamma_{10}$.

To construct the Yukawa terms, one also introduces the following fermionic field:
\begin{equation}
\tilde{\psi}^{(16)\,f}_\alpha= ({\psi}^{(16)\,f}_\alpha)^{\top}B,\quad B=\prod_{i=odd}\Gamma^i.
\label{tildefield}
\end{equation}
Taking into account that $(\Sigma^{ij})^{\top}B=-B\Sigma^{ij}$, one easily deduces that the BRS transformation of
$\tilde{\psi}^{(16)\,f}_\alpha$ is given by
\begin{equation}
s\tilde{\psi}^{(16)\,f}_\alpha=i\tilde{\psi}^{(16)\,f}_\alpha\,c.
\label{BRSoftilde}
\end{equation}

\section{Introducing the noncommutative fields of the noncommutative SO(10) GUT.}

Within the enveloping-algebra framework of Refs.~\cite{Madore:2000en, Jurco:2000ja, Jurco:2001rq}, one introduces
at least a noncommutative field for each ordinary field. Each noncommutative field  is a function --called the Seiberg-Witen map-- of its ordinary counterpart, the ordinary gauge field and the noncommutativity matrix $\omega^{\mu\nu}$. This function --ie, the Seiberg-Wittem map-- maps infinitesimal gauge orbits of the ordinary fields into noncommutative gauge orbits of their noncommutative counterparts. We shall  assume --as suits the Feynman-diagram language-- that the Seiberg-Witten map in momentum space is given by a formal power series expansion in the ordinary fields.

Let us first introduce the noncommutative gauge field, which we shall denote by $A_{\mu}[a_\nu;\omega]$, which is the counterpart of the ordinary field $a_\mu$ in ~(\ref{ordsvec}). $A_{\mu}[a_\nu;\omega]$ is a solution to the following set of Seiberg-Witten map equations:
\begin{equation}
\begin{array}{l}
{s_{nc}C[a_\mu,c;\omega]=s C[a_\mu,c;\omega],\quad
s_{nc}A_{\mu}[a_\nu;\omega]=sA_{\mu}[a_\nu;\omega],}\\[4pt]
{C[a_\mu,c;\omega=0]=c,\quad A_{\mu}[a_\nu;\omega=0]=a_{\mu},}\\[4pt]
{(C[a_\mu,c;\omega])^{\dagger}=C[a_\mu,c;\omega],\quad (A_{\mu}[a_\nu;\omega])^{\dagger}=A_{\mu}[a_\nu;\omega],}
\label{ASWmape}
\end{array}
\end{equation}
where $C[a_\mu,c;\omega]$ is the noncommutative ghost field, $s$ is the ordinary BRS operator in ~(\ref{ordBRStrans}) and $s_{nc}$ is the noncommutative BRS operator defined as follows
\begin{equation}
s_{nc}C=-iC\star C,\quad  s_{nc}A_{\mu}=\partial_{\mu}C+i[A_{\mu},C]_{\star}.
\label{Ancbrstrans}
\end{equation}
Here, $C=C[a_\mu,c;\omega]$ and $A_{\mu}=A_{\mu}[a_\nu;\omega]$. The noncommutative field $A_{\mu}[a_\nu;\omega]$  is an element of the universal enveloping algebra of the Lie algebra of S0(10) in the representation induced by the Dirac matrices $\Gamma^{i}$, $i=1...10$; $A_{\mu}[a_\nu;\omega]$ is, therefore, an element of $\mathbb{C}\rm{l}_{10}(\mathbb{C})$.

Next, $a^{(H)}_{\mu}$ in ~(\ref{hafield}) gives rise to a noncommutative gauge field, which we shall denote by $A^{(H)}_{\mu}[a^{(H)}_\nu;\omega]$. $A^{(H)}_{\mu}[a^{(H)}_\nu;\omega]$ solves
the following set of Seiberg-Witten map equations
\begin{equation}
\begin{array}{l}
{s_{nc}C^{(H)}[a^{(H)}_\mu,c^{(H)};\omega]=s C^{(H)}[a^{(H)}_\mu,c^{(H)};\omega],\quad
s_{nc}A^{(H)}_{\mu}[a^{(H)}_\nu;\omega]=sA^{(H)}_{\mu}[a^{(H)}_\nu;\omega],}\\[4pt]
{C^{(H)}[a^{(H)}_\mu,c^{(H)};\omega=0]=c^{(H)},\quad A^{(H)}_{\mu}[a^{(H)}_\nu;\omega=0]=a^{(H)}_{\mu},}\\[4pt]
{(C^{(H)}[a^{(H)}_\mu,c^{(H)};\omega])^{\dagger}=C^{(H)}[a^{(H)}_\mu,c^{(H)};\omega],\quad (A^{(H)}_{\mu}[a^{(H)}_\nu;\omega])^{\dagger}=A^{(H)}_{\mu}[a^{(H)}_\nu;\omega],}
\label{HASWmape}
\end{array}
\end{equation}
where $C^{(H)}[a^{(H)}_\mu,c^{(H)};\omega]$ is the noncommutative ghost field and $s_{nc}$ is the noncommutative BRS operator defined in ~(\ref{Ancbrstrans}), but now $C=C^{(H)}[a^{(H)}_\mu,c;\omega]$ and $A_{\mu}=A^{(H)}_{\mu}[a^{(H)}_\nu;\omega]$. The noncommutative field $A^{(H)}_{\mu}[a^{(H)}_\nu;\omega]$  is an element of the universal enveloping algebra of the Lie algebra of S0(10) in the representation induced the representation carried by $\varphi^{(H)}$. Recall that $H$ labels the representation and that $H= 210, 10, 45, \overline{126}$ and $120$.

We shall need the noncommutative field strengths, $F_{\mu\nu}[a_\mu;\theta]$ and $F^{(H)}_{\mu\nu}[a_\mu;\theta]$,  to define the Yang-Mills action on noncommutative space-time for our noncommutative S0(10) GUT. We define
\begin{equation}
\begin{array}{l}
{F_{\mu\nu}[a_\rho;\theta]=\partial_{\mu}A_{\nu}-\partial_{\nu}A_{\mu}+i[A_{\mu},A_{\nu}]_{\star},\quad A_{\mu}=A_{\mu}[a_\nu;\omega],}\\[4pt]
{F^{(H)}_{\mu\nu}[a^{(H)}_\rho;\theta]=\partial_{\mu}A_{\nu}-\partial_{\nu}A_{\mu}+i[A_{\mu},A_{\nu}]_{\star},\quad A_{\mu}=A^{(H)}_{\mu}[a^{(H)}_\nu;\omega].}
\end{array}
\label{ncfieldstrengh}
\end{equation}
Notice that $F_{\mu\nu}[a_\rho;\theta]$ belongs to $\mathbb{C}\rm{l}_{10}(\mathbb{C})$ and $F^{(H)}_{\mu\nu}[a_\rho;\theta]$ takes values in the universal enveloping algebra of de Lie algebra of SO(10) in the representation induced by the representation $H$ of the latter.

Using~(\ref{Ancbrstrans}), one can show that
\begin{equation}
sF_{\mu\nu}[a_\rho;\theta]=i[F_{\mu\nu}[a_\rho;\theta], C]_{\star}=s_{nc}F_{\mu\nu}[a_\rho;\theta],
\label{FBRStrans}
\end{equation}
if $C=C[a_\mu,c;\omega]$. The same equation holds for $F^{(H)}_{\mu\nu}[a^{(H)}_\rho;\theta]$, {\it mutatis mutandis}.

The noncommutative fermionic fields will be denoted by $\Psi^{(16)\, f}_\alpha[a_\mu,\psi^{(16)\, f}_\alpha;\theta]$ and $\tilde{\Psi}^{(16)\, f}_\alpha[a_\mu,\tilde{\psi}^{(16)\, f}_\alpha;\omega]$. These noncommutative fermonic fields satisfy the following equations:
\begin{equation}
\begin{array}{l}
{s_{nc}\Psi^{(16)\,f}_\alpha[a_\mu,\psi^{(16)\,f}_\alpha;\omega]=s\Psi^{(16)\,f}_\alpha[a_\mu,\psi^{(16)\,f}_\alpha;\theta],\quad \Psi^{(16)\,f}_\alpha[a_\mu,\tilde{\psi}^{(16)\,f}_\alpha;\omega=0]=\psi^{(16)\,f}_\alpha}\\[4pt]
{s_{nc}\tilde{\Psi}^{(16)\,f}_\alpha[a_\mu,\tilde{\psi}^{(16)\,f}_\alpha;\omega]=
s\tilde{\Psi}^{(16)\,f}_\alpha[a_\mu,\tilde{\psi}^{(16)\,f}_\alpha;\omega],\quad
\tilde{\Psi}^{(16)\,f}_\alpha[a_\mu,\tilde{\psi}^{(16)\,f}_\alpha;\omega=0]=\tilde{\psi}^{(16)\,f}_\alpha}.
\end{array}
\label{psiSWmape}
\end{equation}
The ordinary BRS operator, $s$, acts on the ordinary fermionic fields as given in ~(\ref{ordBRStrans}) and ~(\ref{BRSoftilde}).

The action of noncommutative BRS operator, $s_{nc}$, on $\Psi^{(16)\, f}_\alpha[a_\mu,\psi^{(16)\, f};\omega]$ and $\tilde{\Psi}^{(16)\, f}_\alpha[a_\mu,\tilde{\psi}^{(16)\, f};\omega]$ is given by
the formulae:
\begin{equation}
s_{nc}\Psi_\alpha=-iC\star\Psi_\alpha,\quad s_{nc}\tilde{\Psi}_\alpha=i\tilde{\Psi}_\alpha\star C,
\label{PsiBRStrans}
\end{equation}
with $C=C[a_\mu,c;\omega]$, $\Psi_\alpha=\Psi^{(16)\, f}_\alpha[a_\mu,\psi^{(16)\, f};\omega]$ and $\tilde{\Psi}_\alpha=\tilde{\Psi}^{(16)\, f}_\alpha[a_\mu,\tilde{\psi}^{(16)\, f};\omega]$.

The noncommutative Higgs field which is the noncommutative counterpart of the ordinary Higgs multiplet $\varphi^{(H)}$, $H=210, 10, 45, \overline{126}, 120$, introduced below ~(\ref{selfdual}),
will be denoted by $\hat{\varphi}^{(H)}[a^{(H)}_\mu,\varphi^{(H)};\omega]$. $\hat{\varphi}^{(H)}[a^{(H)}_\mu,\varphi^{(H)};\omega]$ solves the following equations:
\begin{equation}
s_{nc}\hat{\varphi}^{(H)}[a^{(H)}_\mu,\varphi^{(H)};\omega]=s\hat{\varphi}^{(H)}[a^{(H)}_\mu,\varphi^{(H)};\omega],\quad \hat{\varphi}^{(H)}[a^{(H)}_\mu,\varphi^{(H)};\omega=0]=\varphi^{(H)},
\label{varPhiSWmap}
\end{equation}
where, by definition,
\begin{equation}
s_{nc}\hat{\varphi}^{(H)}[a^{(H)}_\mu,\varphi^{(H)};\omega]=-iC \star hat{\varphi}^{(H)}[a^{(H)}_\mu,\varphi^{(H)};\omega],\quad C=C^{(H)}[a^{(H)}_\mu,c^{(H)};\omega].
\label{varphiBRStrans}
\end{equation}

Next, $\Phi^{(H)}[a_\mu,\phi^{(H)};\omega]$, $H=210,
10, 45, \overline{126}, 120$ will stand for the noncommutative counterparts of the ordinary Higgs fields $\phi^{(H)}$ in ~(\ref{Clifhiggs}). The Seiberg-Witten map equations
that solve  $\Phi^{(H)}[a_\mu,\phi^{(H)};\omega]$ read
\begin{equation}
s_{nc}\Phi^{(H)}[a_\mu,\phi^{(H)};\omega]=s\Phi^{(H)}[a_\mu,\phi^{(H)};\omega],\quad \Phi^{(H)}[a_\mu,\phi^{(H)};\omega=0]=\phi^{(H)},
\label{PhiSWmape}
\end{equation}
where $s_{nc}$ is given by
\begin{equation}
s_{nc}\Phi^{(H)}[a_\mu,\phi^{(H)};\omega]=i[\Phi^{(H)}[a_\mu,\phi^{(H)};\omega],C]_{\star},\quad C=C[a_\mu,c;\omega].
\label{PhiBRStrans}
\end{equation}

We shall need later the noncommutative covariant derivatives  of the noncommutative matter fields,  which are given by
\begin{equation}
\begin{array}{l}
{D_{\mu}[A]\Psi^{(16)\, f}_\alpha=\partial_{\mu}\Psi^{(16)\, f}_\alpha+iA_{\mu}\star\Psi^{(16)\, f}_\alpha,\quad
D_{\mu}[A]\tilde{\Psi}^{(16)\, f}_\alpha=\partial_{\mu}\tilde{\Psi}^{(16)\, f}_\alpha-i \tilde{\Psi}^{(16)\, f}_\alpha\star A_{\mu},}\\[4pt]
{D_{\mu}[A^{(H)}]\hat{\varphi}^{(H)}=\partial_{\mu}\hat{\varphi}^{(H)}+i A^{(H)}_{\mu}\star\hat{\varphi}^{(H)},\quad
D_{\mu}[A^{(H)}]\Phi^{(H)}=\partial_{\mu}\Phi^{(H)}+i[A_{\mu},\Phi^{(H)}]_{\star},}
\end{array}
\label{nccovder}
\end{equation}
with $\Psi^{(16)\, f}_\alpha=\Psi^{(16)\, f}_\alpha[a_\mu,\psi^{(16)\, f};\omega]$, $\tilde{\Psi}^{(16)\, f}_\alpha=\tilde{\Psi}^{(16)\, f}_\alpha[a_\mu,\tilde{\psi}^{(16)\, f};\omega]$,
$\hat{\varphi}^{(H)}=\hat{\varphi}^{(H)}[a^{(H)}_\mu,\varphi^{(H)};\omega]$,
$\Phi^{(H)}=\Phi^{(H)}[a_\mu,\phi^{(H)};\omega]$,  $A^{(H)}_\mu=A^{(H))}_{\mu}[a^{(H)}_\nu;\omega]$ and    $A_\mu=A_{\mu}[a_\nu;\omega]$.

\section{The Peccei-Quinn charges of the noncommutative fields}

The need  to solve the strong CP problem and to explain why  Dark Matter exists in the observed amount demands ~\cite{Altarelli:2013aqa} that the ordinary SO(10) GUT of Ref.~\cite{Altarelli:2013aqa}  should have a spontaneously broken Peccei-Quinn symmetry --
see also Refs.~\cite{Holman:1982tb, Bajc:2005zf}. The particle, called the axion, this spontaneously broken global symmetry gives rise to may constitute the Dark Matter of the Universe.

The Peccei-Quinn symmetry is a global U(1) symmetry of the action of the ordinary theory. The invariance of the ordinary Yukawa terms under the Peccei-Quinn symmetry imposes the following transformation
laws on the ordinary fields $\psi^{(16)}_{\alpha}$, $\phi^{(10)}$, $\phi^{(\overline{126})}$ and $\phi{(120)}$:
\begin{equation*}
\psi^{(16)}_{\alpha}\rightarrow e^{iQ\theta}\,\psi^{(16)}_{\theta},\quad \phi^{(10)}\rightarrow e^{-i2Q\theta}\phi^{(10)},\quad \phi^{(\overline{126})}\rightarrow e^{-i2Q\theta}\phi^{(\overline{126})},\quad
\phi^{(120)}\rightarrow e^{-i2Q\theta}\phi^{(120)},
\end{equation*}
where we have chosen the Peccei-Quinn charge, $Q$, of the fermionic multiplet as the unit of the Peccei-Quinn charge. Notice that the Higgs fields  $\phi^{(10)}$,  $\phi^{(120)}$ must chosen to be non-hermitian,  thus giving rise to two irreps of SO(10) each. In Ref.~\cite{Altarelli:2013aqa}, it has been shown that $\phi^{(45)}$ cannot be neutral under the Peccei-Quinn U(1), but with charge
$Q'\neq Q$:
\begin{equation*}
\phi^{(45)}\rightarrow e^{iQ'\theta}\phi^{(45)}.
\end{equation*}
All the other fields of the SO(10) GUT are chosen to be neutral under the Peccei-Quinn symmetry.

We shall be conservative and impose  that global symmetries of the action are not modified by the noncommutative character of space-time. Hence, we shall choose Seiberg-Witten maps such that
the Peccei-Quinn charge of each noncommutative field is well-defined and agrees with that of its ordinary counterpart, ie,
\begin{equation}
\begin{array}{l}
{\Psi^{(16)}_{\alpha}[a_\mu, e^{iQ\theta}\psi^{(16)}_\alpha;\omega]=e^{iQ\alpha}\,\Psi^{(16)}_{\alpha}[a_\mu,\psi^{(16)}_\alpha ;\omega],\;
\Phi^{(10)}[a_\mu, e^{-i2Q\theta}\phi^{(10)};\omega]=e^{-i2Q\theta}\,\Phi^{(10)}[a_\mu,\phi^{(10)} ;\omega],}\\[4pt]
{\Phi^{(\overline{126})}[a_\mu, e^{-i2Q\theta}\phi^{(\overline{120})};\omega]=e^{-i2Q\theta}\,\Phi^{(\overline{126})}[a_\mu,\phi^{(\overline{126})} ;\omega],}\\[4pt]
{\Phi^{(120)}[a_\mu, e^{-i2Q\theta}\phi^{(120)};\omega]=e^{-i2Q\alpha}\,\Phi^{(120)}[a_\mu,\phi^{(120)} ;\omega],}\\[4pt]
{\Phi^{(45)}[a_\mu, e^{iQ'\theta}\phi^{(45)};\omega]=e^{i2Q'\alpha}\,\Phi^{(45)}[a_\mu,\phi^{(45)} ;\omega],}\\[4pt]
{\hat{\varphi}^{(10)}[a^{(10)}_\mu, e^{-i2Q\theta}\varphi^{(10)};\omega]=e^{-i2Q\theta}\,\hat{\varphi}^{(10)}[a^{(10)}_\mu,\varphi^{(10)} ;\omega],}\\[4pt]
{\hat{\varphi}^{(\overline{126})}[a^{(\overline{126})}_\mu, e^{-i2Q\theta}\varphi^{(\overline{120})};\omega]=
e^{-i2Q\theta}\,\hat{\varphi}^{(\overline{126})}[a^{(\overline{126})}_\mu,\varphi^{(\overline{126})} ;\omega],}\\[4pt]
{\hat{\varphi}^{(120)}[a^{(120)}_\mu, e^{-i2Q\theta}\varphi^{(120)};\omega]=e^{-i2Q\theta}\,\hat{\varphi}^{(120)}[a^{(120)}_\mu,\varphi^{(120)} ;\omega],}\\[4pt]
{\hat{\varphi}^{(45)}[a^{(45)}_\mu, e^{iQ'\theta}\varphi^{(45)};\omega]=e^{i2Q'\theta}\,\hat{\varphi}^{(45)}[a^{(45)}_\mu,\varphi^{(45)} ;\omega].}
\end{array}
\label{ncpeccei}
\end{equation}
 That there are Seiberg-Witten maps satisfying the transformation laws in ~(\ref{ncpeccei}) is a consequence of the fact that the Seiberg-Witten map for the matter fields can always be chosen so that
it is linear in the corresponding ordinary fields. See  Ref.~\cite{Martin:2012aw} and the Appendix, for further details.

\section{The action of the noncommutative SO(10) GUT.}

Let us first point out that we shall assume that Lorentz indices are raised and lowered
with the help of the Minkowski metric $(-,+,+,+)$.

The action, $S$,  which gives the dynamics of our noncommutative GUT, will be sum of integrated  monomials with regard to the $\star$-product of the noncommutative fields, introduced in the previous section, and their derivatives. We shall restrict the mass dimension of these monomials to be less than or equal to 4, since  we are interested in constructing the noncommutative counterpart of a renormalizable ordinary SO(10) GUT. So, not considering monomials with mass dimension bigger than 4 is the simplest choice to start with. For the sake of simplicity, We shall also assume that the dependence on $\omega^{\mu\nu}$ of the noncommutative action only occurs through the Seiberg-Witten map and the $\star$-product. We shall demand that the noncommutative action be invariant under the noncommutative
BRS transformations defined in ~(\ref{Ancbrstrans}), ~(\ref{PsiBRStrans}) and ~(\ref{PhiBRStrans}) and the Peccei-Quinn transformations in  ~(\ref{ncpeccei}). We shall break the action
into four parts:
\begin{equation}
S=S_{YM}+S_{fermionic}+S_{Yukawa}+S_{Higgs}
\label{NCaction}
\end{equation}
and discuss each part separately below.

\subsection{The noncommutative Yang-Mills action}

In view of ~(\ref{FBRStrans}) and following Ref.~\cite{Aschieri:2002mc}, we shall define the noncommutative Yang-Mills action, $S_{YM}$, as follows
\begin{equation}
S_{YM}\,=\, -\kappa_c\idx\, \Tr\, F_{\mu\nu}[a_\rho;\omega]\star F^{\mu\nu}[a_\rho;\omega]\,-\,\sum_{H}\kappa^{(H)}\,\idx\,\Tr\, F^{(H)}_{\mu\nu}[a^{(H)}_\rho;\omega]\star F^{(H)\,\mu\nu}[a^{(H)}_\rho;\omega].
\label{deffsym}
\end{equation}
$F_{\mu\nu}[a_\rho;\omega]$ and $F^{(H)}_{\mu\nu}[a^{(H)}_\rho;\omega]$ are given in ~(\ref{ncfieldstrengh}) and the real constants $\kappa_c$, $\kappa_{H}$ are constrained by the following equation:
 \begin{equation}
\frac{1}{g_{YM}^2}\,=\,32\,k_c\,+\, \sum_{H}\,4\,I_2(H)\,\kappa_{H}.
\label{constraint}
\end{equation}
$g_{YM}$ is the tree-level Yang-Mills coupling constant and $I_2(H)$ is the second order Dynkin index of the irrep $H$ of  SO(10), ie,
\begin{equation*}
\Tr\; M^{(H)}_{ij}\,M^{(H)}_{kl}\,=\,I_2(H)(\delta_{ik}\delta_{jl}-\delta_{il}\delta_{jk}).
\end{equation*}
Notice that ~(\ref{constraint}) comes from the need to have the right normalization of the free kinetic term of the gauge field. Positivity of $S_{YM}$ for Euclidean signature puts further constraints on the $\kappa$'s, which are automatically satisfied if  the $\kappa$'s are positive or vanish.

We should like to point out that there is no reason to set to zero from the beginning any of the  $\kappa$'s in ~(\ref{deffsym}), for one-loop  Higgs radiative corrections  generate contributions --see Ref.~\cite{Martin:2007wv}-- to the effective action of gauge field that are of the type:
\begin{equation*}
\idx\,\Tr\, F^{(H)}_{\mu\nu}[a^{(H)}_\rho;\omega]\star F^{(H)\,\mu\nu}[a^{(H)}_\rho;\omega].
\end{equation*}

\subsection{The fermionic part of action}

Furnished with the noncommutative fermionic fields defined by ~(\ref{psiSWmape}) and their covariant derivatives in ~(\ref{nccovder}), one constructs the fermionic part of the action, $S_{fermionic}$, of our noncommutative GUT.
We shall assume that $S_{fermionic}$ is quadratic in the noncommutative fermionic fields and linear in the noncommutative gauge field. This is the simplest choice for an
$S_{fermionic}$. $S_{fermionic}$ reads
\begin{equation*}
S_{fermionic}=\sum_{f} \Big(\kappa^{f}\idx\,i\Psi^{f\,\dagger}_{\dot{\alpha}}\bar{\sigma}^{\dot{\alpha}\alpha}_{\mu}D^{\mu}[A]\Psi^{f}_{\alpha}-
\tilde{\kappa}^{f}\idx\,iD_{\mu}[A]\tilde{\Psi}^{f\,\alpha}\sigma^{\mu}_{\alpha\dot{\alpha}}\tilde{\Psi}^{f\,\dagger\,\dot{\alpha}}\Big),
\end{equation*}
where $\Psi^{f}_{\alpha}=\Psi^{(16)\, f}_\alpha[a_\mu,\psi^{(16)\, f}_\alpha;\omega]$, $\tilde{\Psi}^{f}_{\alpha}=\tilde{\Psi}^{(16)\, f}_\alpha[a_\mu,\tilde{\psi}^{(16)\, f}_\alpha;\omega]$, $\Psi^{f\,\dagger}_{\dot{\alpha}}=(\Psi^{f}_{\alpha})^{\dagger}$ and
$\tilde{\Psi}^{f\,\dagger}_{\dot{\alpha}}=(\tilde{\Psi}^{f}_{\alpha})^{\dagger}$.
 The conventions on dotted and undotted indices that we use are those of Ref.~\cite{Martin:2012us}. The proper normalization of the free propagator of the fermions demands that
 \begin{equation*}
 \kappa^{f}+\tilde{\kappa}^{f}=1.
 \end{equation*}

\subsection{The Yukawa terms}

Let us recall --see, for instance, Ref.~\cite{Bajc:2005zf}-- that the most general Yukawa terms in a renormalizable ordinary SO(10) GUT with only fermionic multiplets in the 16 irrep reads
\begin{equation*}
\sum_{ff'}\idx\; \Big( Y^{(10)}_{ff'}\,\tilde{\psi}^{(16)\,f\,\alpha}\phi^{(10)}\psi^{(16)\,f'}_{\alpha}+
Y^{(\overline{126})}_{ff'}\,\tilde{\psi}^{(16)\,f\,\alpha}\phi^{(\overline{126})}\psi^{(16)\,f'}_{\alpha}+
Y^{(120)}_{ff'}\,\tilde{\psi}^{(16)\,f\,\alpha}\phi^{(120)}\psi^{(16)\,f'}_{\alpha}+h.c.\Big).
\end{equation*}
Recall that $\phi^{(10)}$, $\phi^{(\overline{126})}$ and $\phi^{(120)}$ have been defined in  ~(\ref{Clifhiggs}) and $\tilde{\psi}^{(16)\,f\,\alpha}$ has been introduced in
~(\ref{tildefield}). It is apparent that the simplicity and beauty of the previous expression comes from the fact that  Higgs fields which occur in it can be interpreted as elements of $\mathbb{C}\rm{l}_{10}(\mathbb{C})$. It is also this feature of the Higgs fields in the ordinary Yukawa terms above that enables us to introduce, in a natural way,  the following noncommutative Yukawa terms:
\begin{equation}
\begin{array}{l}
{S_{Yukawa}=
\sum_{ff'}\idx\; \Big( Y^{(10)}_{ff'}\,\tilde{\Psi}^{(16)\,f\,\alpha}\star\Phi^{(10)}\star\Psi^{(16)\,f'}_{\alpha}+
Y^{(\overline{126})}_{ff'}\,\tilde{\Psi}^{(16)\,f\,\alpha}\star\Phi^{(\overline{126})}\star\Psi^{(16)\,f'}_{\alpha}}\\[4pt]
{\phantom{S_{Yukawa}=
\sum_{ff'}\idx\quad }+Y^{(120)}_{ff'}\,\tilde{\Psi}^{(16)\,f\,\alpha}\star\Phi^{(120)}\star\Psi^{(16)\,f'}_{\alpha}+h.c.\Big)},
\end{array}
\label{ncyukawas}
\end{equation}
where
\begin{equation*}
\begin{array}{l}
{\tilde{\Psi}^{(16)\,f}_{\alpha}=\tilde{\Psi}^{(16)\,f}_{\alpha}[a_\mu,\tilde{\Psi}^{(16)\,f}_\alpha;\omega],\quad
\Psi^{(16)\,f}_{\alpha}=\Psi^{(16)\,f}_{\alpha}[a_\mu,\Psi^{(16)\,f}_\alpha;\omega],}\\[4pt]
{\Phi^{(H)}=\Phi^{(H)}[a^{(H)}_\mu,\phi^{(H)};\omega],\quad H=10, \overline{126}, 120.}
\end{array}
\end{equation*}
are Seiberg-Witten maps that solve ~(\ref{psiSWmape}) and ~(\ref{PhiSWmape}). By construction, $S_{Yukawa}$, in ~(\ref{ncyukawas}), is invariant under BRS transformations of the ordinary fields and the corresponding noncommutative BRS transformations in ~(\ref{PsiBRStrans}) and ~(\ref{PhiBRStrans}).

Let us point out that the noncommutative Higgses $\hat{\varphi}^{(H)}_{I}$, $I=1...\,dim\, H$, $H=10,\overline{126}, 120$ defined by ~(\ref{varPhiSWmap}) are of no help for constructing integrated cubic monomials, with regard to the $\star$-product, of the noncommutative fields $\tilde{\Psi}^{(16)\,f}_{\alpha}$, $\Psi^{(16)\,f}_{\alpha}$ and $\hat{\varphi}^{(H)}_{I}$. Indeed, let
$\mathscr{T}^{rIr'}$ be complex numbers, then, the fact that the $\star$-product
is not commutative and that $C[a_\mu,c;\omega]$ and $C^{(H)}[a^{(H)}_\mu,c^{(H)};\omega]$ take values in the enveloping algebra of SO(10), prevents  the term
\begin{equation}
\idx\; \mathscr{T}^{rIr'}\; \tilde{\Psi}^{(16)\,f\,\alpha}_r\star\hat{\varphi}^{(H)}_{I}\star\Psi^{(16)\,f}_{\alpha\,r'}
\label{nonBRSinv}
\end{equation}
from being invariant under the noncommutative BRS transformations in ~(\ref{PsiBRStrans}) and ~(\ref{varphiBRStrans}). This result holds whatever the ordering of the fields in ~(\ref{nonBRSinv}).

In view of the previous discussion the reader may ask why we have introduced the noncommutative Higgs fields $\hat{\varphi}^{(H)}_{I}$. We shall answer this question in the next subsection.

\subsection{The Higgs action}

$S_{Higgs}$ in ~(\ref{NCaction}) contains only Higgs fields and gauge fields. Let us begin  by introducing the kinetic terms. These we take to be quadratic in the noncommutative Higgs fields and their covariant derivatives. We also assume that --as in the ordinary field theory case-- these terms are semi-positive definite after a Wick rotation has been performed. We thus end up with the following gauge covariant kinetic terms for the noncommutative Higgses:
\begin{equation}
\mathcal{K}_{Higgs}=-\sum_{H}\,\idx\; \Big(s_{H}\, (D_{\mu}[A^{(H)}]\hat{\varphi}^{(H)})^{\dagger}\,D_{\mu}[A^{(H)}]\hat{\varphi}^{(H)}+t_H\,\Tr\,\big((D_{\mu}[A]\Phi^{(H)})^{\dagger}\,D_{\mu}[A]\Phi^{(H)}\big)\Big),
\label{Kinetic}
\end{equation}
where $D_{\mu}[A^{(H)}]\hat{\varphi}^{(H)}$ and $D_{\mu}[A]\Phi^{(H)}$ are given in ~(\ref{nccovder})  and $H=210, 10, 45,  \overline{126}$ and $120$. The parameters $s_H$ and $t_H$ are positive real numbers such that  each free kinetic term has the right normalization.

It is plain that the second summand of ~(\ref{Kinetic}) is needed, for the noncommutative Yukawa terms in ~(\ref{ncyukawas}) involve $\Phi^{(H)}$. To provide the rationale for the first summand of
~(\ref{Kinetic}), we must discuss why the construction of a phenomenologically  sensible  noncommutative Higgs potential seems to require that the noncommutative fields $\hat{\varphi}$ be added to the pool of noncommutative Higgs fields.

In Ref.~\cite{Bertolini:2009es}, it has been analyzed the classical vacuum structure of an ordinary S0(10) GUT with a  Higgs in the 45 and another in the 16. It has been shown there that if the monomial
$\Tr((\phi^{(45)})^{\dagger}\phi^{(45)}(\phi^{(45)})^{\dagger}\phi^{45)})$ occurs in the Higgs potential, then the monomial $(\Tr((\phi^{(45)})^{\dagger}\phi^{(45)}))^2$ must be also a summand of the Higgs potential. This result comes from  demanding boundedness from below of the Higgs potential and absence of tachyons.

Now, it is true that the ordinary GUT corresponding to our noncommutative GUT has a Higgs content more involved  than the one used in Ref.~\cite{Bertolini:2009es} and that no analysis similar to that in the latter paper has been carried out for an ordinary GUT with our Higgs content; so it cannot be claimed that both type of monomials must necessarily occur in the Higgs potential. However,  until such an complicated analysis is carried out in the ordinary case, we  shall  play it safe and include  in the noncommutative Higgs potential noncommutative counterparts of both $\Tr((\phi^{(H)})^{\dagger}\phi^{(H)}(\phi^{(H)})^{\dagger}\phi^{H)})$ and $(\Tr((\phi^{(H)})^{\dagger}\phi^{(H)}))^2$, $H=210, 10, 45, \overline{126}$ and $120$.

It is plane that
\begin{equation*}
\idx\;\Tr\;(\Phi^{(H)})^{\dagger}\star\Phi^{(H)}\star(\Phi^{(H)})^{\dagger}\star\Phi^{H)}))\quad\text{and}\quad
\idx\;\Tr\;(\Phi^{(H)})^{\dagger}\star(\Phi^{(H)})^{\dagger}\star\Phi^{(H)}\star(\Phi^{(H)}),
\end{equation*}
are invariant under the noncommutative BRS transformations in ~(\ref{PhiBRStrans}), provided $\Phi^{(H)}$ is given by a Seiberg-Witten map. However,
\begin{equation*}
\idx\;(\Tr\;((\Phi^{(H)})^{\dagger}\star\Phi^{(H)}))^2
\end{equation*}
is not invariant under the noncommutative BRS transformations in ~(\ref{PhiBRStrans}), for --unlike the ordinary case-- the unintegrated monomial $(\Tr\;((\Phi^{(H)})^{\dagger}\star\Phi^{(H)}))^2)$ is not invariant under  BRS transformations and, what is worse,  its BRS variation is not  a total derivative. Hence, to construct the noncommutative counterpart of the ordinary integrated monomial
\begin{equation}
\idx\;(\Tr\;((\phi^{(H)})^{\dagger}\phi^{(H)}))^2,
\label{squarephi}
\end{equation}
we shall take into account that $\Tr\;((\phi^{(H)})^{\dagger}\phi^{(H)}))=s_H\,(\varphi^{(H)})^{\dagger\, I}\varphi_I^{(H)})$, where $s_H$ is a real number --$s_{210}=32\times 4!, s_{10}= 32,
s_{45}=16,...$-- whose actual value is irrelevant to our discussion, and use the noncommutative scalar $\hat{\varphi}_I^{(H)}$ in ~(\ref{varPhiSWmap}), rather than $\Phi^{(H)}$, to define the noncommutative counterpart of the integrated ordinary monomial in ~(\ref{squarephi}) as follows
\begin{equation*}
\idx\;((\hat{\varphi}^{(H)})^{\dagger\,I}\star\hat{\varphi}_I^{(H)}))^2.
\end{equation*}

Let $F^{(H)}$ stand for either $\Phi^{(H)}$ or its hermitian conjugate $(\Phi^{(H)})^{\dagger}$. Then, We are now ready to introduce the Higgs potential, $V_{Higgs}$, of our noncommutative
SO(10) GUT:
\begin{equation}
\begin{array}{l}
{V_{Higgs}=\sum_{H} \alpha_H\,(\hat{\varphi}^{(H)})^{\dagger\,I}\star\hat{\varphi}_I^{(H)}+
\sum_{H_1 H_2} \beta_{H_1,H_2}\,((\hat{\varphi}^{(H_1)})^{\dagger\,I}\star\hat{\varphi}_I^{(H_1)})\star
((\hat{\varphi}^{(H_2)})^{\dagger\,I}\star\hat{\varphi}_I^{(H_2)})\,+\,}\\[4pt]
{\phantom{V_{Higgs}=\,}\gamma_{210}\,\Tr\,\Phi^{(210)}\,+\,}\\[4pt]
{\phantom{V_{Higgs}=\,}\sum_{H_1 H_2}\kappa_{H_1 H_2}\,\Tr\, (F^{(H_1)}\star F^{(H_2)})\,+\,\sum_{H_1 H_2 H_3}\,\gamma_{H_1 H_2 H_3}\Tr\, (F^{(H_1)}\star F^{(H_2)}\star F^{(H_3)})\,+}\\[4pt]
{\phantom{V_{Higgs}=\,}\sum_{H_1 H_2 H_3 H_4}\lambda_{H_1 H_2 H_3 H_4}\,\Tr\, (F^{(H_1)}\star F^{(H_2)}\star F^{(H_3)}\star F^{(H_4)}), }
\end{array}
\label{Higgspotential}
\end{equation}
where $\alpha_H$, $\beta_{H_1 H_2}$, $\kappa_{H_1 H_2}$,  $\gamma_{H_1 H_2 H_3}$ $\lambda_{H_1 H_2 H_3 H_4}$ are numbers; which are real, if the  monomial they go with is real. Boundedness from below of $V_{Higgs}$ put constraints
on the couplings $\beta_H$'s and $\lambda_{H_1 H_2 H_3 H_4}$'s. $H$, $H_1$, $H_2$, $H_3$ and $H_4$ run over the set $210, 10, 45, \overline{126}, 126, 120$.
The monomials  $\Tr\, (F^{(H_1)}\star F^{(H_2)})$, $\Tr\,(F^{(H_1)}\star F^{(H_2)}\star F^{(H_3)})$ and  $\Tr\, (F^{(H_1)}\star F^{(H_2)}\star F^{(H_3)}\star F^{(H_4)})$ satisfy the following conditions:
${\it i}$) The sum of the Peccei-Quinn charges of the fields entering the monomial must vanish --thus, the action will have a Peccei-Quinn symmetry, ${\it ii}$) if a monomial occurs, so does its hermitian conjugate multiplied by the appropriate complex conjugate coupling constant,
and ${\it iii}$) monomials obtained by cyclic permutations of the fields of a given monomial are dropped. Finally, the BRS invariant noncommutative Higgs action, $S_{Higgs}$, reads
\begin{equation*}
S_{Higgs}\,=\,\mathcal{K}_{Higgs}\,-\,\idx\;V_{Higgs},
\end{equation*}
where $\mathcal{K}_{Higgs}$ and $V_{Higgs}$ are given in ~(\ref{Kinetic}) and ~(\ref{Higgspotential}), respectively.

Before closing this section we would like to point out that there monomials in $V_{Higgs}$ that vanish --due to SO(10) invariance-- at $\omega^{\mu\nu}=0$, but are non-vanishing otherwise. This monomials give
rise to intrinsically noncommutative --since they vanish at $\omega^{\mu\nu}=0$-- interactions between  Higgses of several species and the gauge field. Let us give just one example:
\begin{equation*}
\begin{array}{l}
{\idx\,\Tr\,(\Phi^{(10)})^{\dagger}\star\Phi^{(10)}\star\Phi^{(210)}\,=\,4!\times 16\times}\\[4pt]
{\int\!\prod_{i=1}^{4}\;\idpi\,(2\pi)^4\,\delta(-p_1+p_2+p_3+p_4)\,(\varphi^{(10)})^{*}_{i_1}(p_1)\,
\varphi^{(10)}_{j_1}(p_2)\,a^{k_1\,k_2}_{\mu_1}(p_3)\,\varphi^{(210)}_{i_1 j_1 k_1 k_2}(p_4)}\\[4pt]
{[-e^{\frac{i}{2}(p_1-p_3)\wedge p_2)}\,\omega^{\mu_1\mu_2}\,p_{1\,\mu_2}\,
\frac{\sin(\frac{1}{2}p_3\wedge p_1)}{p_3\wedge p_1}\,
+\,e^{\frac{i}{2}p_1\wedge (p_2+p_3)}\,\omega^{\mu_1\mu_2}\,p_{2\,\mu_2}
\frac{\sin(\frac{1}{2}p_3\wedge p_2)}{p_3\wedge p_2}\,+}\\[4pt]
{\phantom{-e^{\frac{i}{2}(p_1-p_3)\wedge p_2)}\omega^{\mu_1\mu_2}\,p_{1\,\mu_2}\,
\frac{\sin(\frac{1}{2}p_3\wedge p_1)}{p_3\wedge p_1}+\,e^{\frac{i}{2}p_1\wedge (p_2+p_3)}\,\omega^{\mu_1\mu_2}\,}
e^{\frac{i}{2}p_1\wedge p_2}\,\omega^{\mu_1\mu_2}\,p_{4\,\mu_2}\,
\frac{\sin(\frac{1}{2}p_3\wedge p_4)}{p_3\wedge p_4}]\,+\,O(a_{\mu}^2)}.
\end{array}
\end{equation*}
 The r.h.s of previous equation has been derived with the help of the results presented in the Appendix. Let us stress that terms like the one in the  previous equation describe the tree-level coupling between different species of ordinary Higgses, and the ordinary gauge field, as they move
around in noncommutative space-time.  Tree-level couplings such as this are not possible in ordinary Minkowski space-time, so its eventual experimental detection will give a clear hint of the noncommutative character of space-time. This strategy to experimentally probe the possible noncommutative character
of space-time was pioneered by the authors of Ref.~\cite{Behr:2002wx}.

Another peculiarity of the noncommutative Higgs potential in ~(\ref{Higgspotential}) is that it contains a term that is proportional to $\Tr\,\Phi^{(210)}$. This term is not forbidden neither by gauge invariance  nor by the Peccei-Quinn
symmetry. Using the results presented in the appendix one shows that $\Tr\,\Phi^{(210)}$ vanishes at $\omega^{\mu\nu}=0$ at that the first non-trivial contribution coming from it occurs at order $(a_{\mu})^2$, thus giving rise to a non-Lorentz invariant coupling between two gauge fields and the 210 Higgs. Notice that similar terms for the other Higgses of the GUT will explicitly break the Peccei-Quinn symmetry and this would not do.

\section{Outlook}

In this paper we have successfully formulated the classical action of a noncommutative SO(10) GUT which is the counterpart of the phenomenologically relevant ordinary SO(10) GUT of Ref. ~\cite{Altarelli:2013aqa}. The next step to take should
be the systematic  study of its properties as a quantum theory. An important issue that should be tackled at once is the analysis of  the UV/IR mixing effects in this noncommutative theory. One may anticipate that these may not generally occur in the same type of terms
as in U(N) theories. Indeed,  for U(N) theories, the UV/IR mixing phenomenon makes the two point function  of the effective action of the ordinary field  develop the  following well-known IR divergences ~\cite{Hayakawa:1999yt}:
\begin{equation*}
                           \Tr\, a_{\mu}(p)\,\frac{\tilde{p}^{\mu}\tilde{p}^{\nu}}{\tilde{p}^4}\,\Tr\, a_{\nu}(-p),\quad
                           \Tr\, a_{\mu}(p)\,\ln(-p^2\tilde{p}^2)(p^2\eta^{\mu\nu}-p^{\mu}p^{\nu})\,\Tr\, a_{\nu}(-p).
\end{equation*}
But these type of terms do not occur for simple gauge groups since now $\Tr\, a_{\mu}=0$. Recall that $\tilde{p}^{\mu}=\omega^{\mu\nu}\,p_{\nu}$.

Another issue that should be addressed is a comprehensive study of the set of classical vacua of our SO(10) GUT and how it is modified at the quantum level. Of course, the phenomenology that the SO(10) GUT presented here gives rise to  should be  studied. In this regard the analysis --perhaps, along the lines of Refs.~\cite{Horvat:2011iv, Horvat:2011bs, Horvat:2011qn, Horvat:2011qg}-- of the neutrino physics that our GUT yields looks particularly interesting.
\newpage

\section{Appendix}

It has been discussed in Ref.~\cite{Martin:2012aw} how to obtain systematically $\theta$-exact solutions to the Seiberg-Witten map equations in ~(\ref{ASWmape}), ~(\ref{HASWmape}), ~(\ref{psiSWmape}) and ~(\ref{varPhiSWmap}). Here, we shall show how to construct a $\theta$-exact a solution to ~(\ref{PhiSWmape}).

Let us first point out that ~(\ref{ASWmape}) holds in any number of space-time dimensions whatever the value of non-commutativity matrix $\omega^{ij}$. Now, assume that we are in 4+1 space-time dimensions and that $\omega^{ij}$, $i,j=0,1,2,3,4$ is such that $\omega^{\mu 4}=0$, $\mu=0,1,2,3$.
It was shown in Refs.~\cite{Barnich:2001mc, Barnich:2003wq} that the following "evolution'' equations give a solution to ~(\ref{ASWmape})
\begin{equation}
\begin{array}{l}
{\frac{d}{dh}C(h\omega)=\frac{1}{4}\omega^{\rho\sigma}\{\partial_{\rho}C(h\omega),A_{\rho}(h\omega)\}_{\star_h},\quad\quad C(h=0)=c,}\\[4pt]
{\frac{d}{dh}A_{\mu}(h\omega)=\frac{1}{2}\omega^{\rho\sigma}\,\{A_{\rho}(h\omega),\partial_{\sigma}A_\mu(h\omega)\}_{\star_h}-\frac{1}{4}\omega^{\rho\sigma}\,\{A_{\rho}(h\omega),\partial_{\mu}A_\sigma(h\omega)\}_{\star_h}+}\\[4pt]
{\phantom{\frac{d}{dh}A_{\mu}[h\omega]=\;\,}\frac{i}{4}\omega^{\rho\sigma}\,\{A_{\rho}(h\omega),[A_\sigma(h\omega),A_\mu(h\omega)]_{\star_h}\}_{\star_h},\quad A_\mu(h=0)= a_\mu,}\\[4pt]
{\frac{d}{dh}A_{4}(h\omega)=\frac{1}{2}\omega^{\rho\sigma}\,\{A_{\rho}(h\omega),\partial_{\sigma}A_4(h\omega)\}_{\star_h}-\frac{1}{4}\omega^{\rho\sigma}\,\{A_{\rho}(h\omega),\partial_{4}A_\sigma(h\omega)\}_{\star_h}+}\\[4pt]
{\phantom{\frac{d}{dh}A_{4}[h\omega]=\;\,}\frac{i}{4}\omega^{\rho\sigma}\,\{A_{\rho}(h\omega),[A_\sigma(h\omega),A_4(h\omega)]_{\star_h}\}_{\star_h},\quad A_4(h=0)=a_4,}
\end{array}
\label{evoleq}
\end{equation}
where the Greek indices run over $0,1,2$ and $3$ and $\star_h$ denotes the Moyal product where $h\omega^{\mu\nu}$ has replaced $\omega^{\mu\nu}$.

Now, neither $A_4$ nor $a_4$ enter the first to equations in ~(\ref{evoleq}), so these two equations gives Seiberg-Witten maps  $C[a_\rho, c;\omega]$ and $A_\mu[a\rho;\omega]$ which do not depend on $a_4$. On the other hand, the last equation in ~(\ref{evoleq}) yields
a Seiberg-Witten map $A_4[a_\mu,a_4;\omega]$ which depends on both $a_\mu$ and $a_4$. Let us particularize ~(\ref{evoleq}) to ordinary fields $a_i=(a_{\mu},a_4)$, $\mu=0,1,2,3$, which do not depend on $x^4$ and ordinary ghost fields which do not depend on $x^4$, either. For these ordinary field configurations
we have that $C[a_\rho, c;\omega]$, $A_\mu[a_\rho;\omega]$ and $A_4[a_\mu,a_4;\omega]$ solving ~(\ref{evoleq}) do not depend on $x^4$, so the are actually noncommutative fields which live in 3+1 space-time dimensions. From this 4-dimensional point of view, $A_\mu[a_\rho;\omega]$ and $C[a_\rho,c;\omega]$ are, respectively, the noncommutative gauge field and the corresponding ghost field --ie, $A_\mu[a_\rho;\omega]$ and $C[a_\rho,c;\omega]$ solve ~(\ref{ASWmape}) in 3+1 dimensions, whereas $A_4[a_\mu,a_4;\omega]$ is a noncommutative field solving
\begin{equation*}
\begin{array}{l}
{sA_{4}[a_\mu, a_4;\omega]=\partial_4 C[a_\mu,c;\omega]+i\Big[[A_{4}[a_\mu, a_4;\omega],C[a_\mu,c;\omega]\Big]_{\star}=-i\Big[C[a_\mu,c;\omega],A_{4}[a_\mu, a_4;\omega]\Big]_{\star},}\\[4pt]
{A_4[\omega=0]=a_4,}
\end{array}
\end{equation*}
since $\partial_4 C[a_\mu,c;\omega]=0$. If, in the previous equation one substitutes $\Phi^{(H)}[a_\rho,\phi^{(H)};\omega]$ for $A_4[a_\rho,\phi^{(H)};\omega]$, one obtains  ~(\ref{PhiSWmape}). Hence, by replacing $A_4$ with $\Phi^{(H)}$ and $a_4$ with $\phi^{(H)}$ in the last equation of
~(\ref{evoleq}), a solution to ~(\ref{PhiSWmape}) will be produced, if the term involving $\partial_4$ is dropped. We have this shown that the "evolution'' problem that yields the Seiberg-Witten map which defines $\Phi^{(H)}$ reads
\begin{equation}
\begin{array}{l}
{\frac{d}{dh}\Phi^{(H)}(h\omega)=\frac{1}{2}\omega^{\rho\sigma}\,\{A_{\rho}(h\omega),\partial_{\sigma}\Phi^{(H)}(h\omega)\}_{\star_h}+\frac{i}{4}\omega^{\rho\sigma}\,\{A_{\rho}(h\omega),[A_\sigma(h\omega),\Phi^{(H)}(h\omega)]_{\star_h}\}_{\star_h},}\\[4pt]
{\Phi^{(H)}(h=0)=\Phi^{(H)}.}
\end{array}
\label{Phievol}
\end{equation}

The $\omega$-exact solution to ~(\ref{Phievol}) that is formal series expansion in power of the ordinary fields is obtained recursively. Let us express $A_\mu[a_\rho;h\omega]$ and $\Phi^{(H)}[a_\rho,\phi^{(H)};h\omega]$ as follows
\begin{equation*}
A_\mu[a_\rho;h\omega]\,=\,\sum_{n>0}\;A^{(n)}_\mu[a_\mu;h\omega],\quad \Phi^{(H)}[a_\rho,\phi^{(H)};h\omega]\,=\,\sum_{n\geq 0}\;\Phi^{(H,n)}[a_\rho,\phi^{(H)};h\omega],
\end{equation*}
where $A^{(n)}_\mu[a_\rho;h\omega]$ and $\Phi^{(H,n)}[a_\rho,\phi^{(H)};h\omega]$ are monomials of degree $n$ with regard to $A_\mu$. Then substituting them in ~(\ref{Phievol}), one obtains the infinite set of equations
\begin{equation}
\begin{array}{l}
{\Phi^{(H,0)}[a_\rho,\phi^{(H)};h\omega]=\phi^{(H)},}\\[4pt]
{\Phi^{(H,1)}[a_\rho,\phi^{(H)};\omega]=\int_{0}^{1}\,dh\,\Big(\frac{1}{2}\omega^{\rho\sigma}\{A^{(1)}_\rho(h\omega),\partial_{\sigma}\phi^{(H)}\}_{\star_h}\Big)}\\[4pt]
{\Phi^{(H,2)}[a_\rho,\phi^{(H)};\omega]=\int_{0}^{1}\,dh\,\Big(\frac{1}{2}\omega^{\rho\sigma}\{A^{(2)}_\rho(h\omega),\partial_{\sigma}\phi^{(H)}\}_{\star_h}+\frac{1}{2}\omega^{\rho\sigma}\{A^{(1)}_\rho(h\omega),\partial_{\sigma}\Phi^{(H,1)}(h\omega)\}_{\star_h}+}\\[4pt]
{\phantom{\Phi^{(H,2)}[a_\rho,\phi^{(H)};\omega]=\int_{0}^{1}\,dh\,\Big(\;}\frac{i}{4}\,\omega^{\rho\sigma}\,\{A^{(1)}_\rho(h\omega),[A^{(1)}_\sigma(h\omega),\phi^{(H)}]_{\star_h}\}_{\star_h}\Big),}\\[4pt]
{\Phi^{(H,n)}[a_\rho,\phi^{(H)};\omega]=\int_{0}^{1}\,dh\,\Big(\sum_{m=0}^{n-1}\,\frac{1}{2}\omega^{\rho\sigma}\{A^{(n-m)}_\rho(h\omega),\partial_{\sigma}\Phi^{(H,m)}(h\omega)\}_{\star_h}+}\\[4pt]
{\phantom{\Phi^{(H,2)}[a_\rho=\int_{0}^{1}\,dh\,\Big(\;}\sum_{m_1+m_2+m_3=n}\frac{i}{4}\,\omega^{\rho\sigma}\,\{A^{(m_1)}_\rho(h\omega),[A^{(m_2)}_\sigma(h\omega),\phi^{(H,m_3)}(h\omega)]_{\star_h}\}_{\star_h}\Big),\quad n\geq 3,}
\end{array}
\label{recursioneq}
\end{equation}
where $m_1>0$, $m_2>0$ and $m_3\geq 0$. We would like to stress that each $\Phi^{(H,n)}[a_\rho,\phi^{(H)};\omega]$ in ~(\ref{recursioneq}) is linear in the ordinary field
$\phi^{(H)}$. Hence, the corresponding equation in ~(\ref{ncpeccei}) holds for this Seiberg-Witten map.

 Next, with the help of  the results presented in Ref.~\cite{Martin:2012aw}, one may workout the r.h.s of each equality in ~(\ref{recursioneq}) recursively. we shall display  below
 the explicit expressions that we have obtained for $\Phi^{(H,1)}[a_\rho,\phi^{(H)};\omega]$ and $\Phi^{(H,2)}[a_\rho,\phi^{(H)};\omega]$:
\begin{equation*}
\begin{array}{l}
{\Phi^{(H,1)}[a_\rho,\phi^{(H)};\omega]=\int\idpone\idptwo\,e^{i(p_1+p_2)x}\;\omega^{\mu_1\mu_2}{p_2}_{\mu_2}  }\\[4pt]
{\phantom{\Phi^{(H,1)}[a_\rho,\phi^{(H)};\omega]=\int\idpone\idptwo e^{i(p_1)x}}
\Big(\frac{e^{\frac{i}{2}p_1\wedge p_2}-1}{p_1\wedge p_2} \phi^{(H)}(p_2) a_{\mu_1}(p_1)-
\frac{e^{-\frac{i}{2}p_1\wedge p_2}-1}{p_1\wedge p_2}  a_{\mu_1}(p_1)\phi^{(H)}(p_2)\Big),}
\end{array}
\end{equation*}
where $p_1\wedge p_2= \omega^{\mu_1\mu_2}\,{p_1}_{\mu_1}{p_2}_{\mu_2}$, and
\begin{equation*}
\begin{array}{l}
{\Phi^{(H,2)}[a_\rho,\phi^{(H)};\omega]=\int\idpone\idptwo\idpthree\,e^{i(p_1+p_2+p_3)x}\;}\\[4pt]
{\Big\{\frac{1}{2}\omega^{\mu\nu}\omega^{\rho\sigma}\,[2({p_2}_{\sigma}\delta^{\mu_1}_\rho\delta^{\mu_2}_\mu+{p_1}_\sigma\delta^{\mu_1}_\mu\delta^{\mu_2}_\rho)
-(p_2-p_1)_\mu\delta^{\mu_1}_\rho\delta^{\mu_2}_\sigma]\,{p_3}_\nu\times}\\[4pt]
{\;\;\big[{\cal G}(-p_3;p_1,p_2;\omega)\,a_{\mu_1}(p_1)a_{\mu_2}(p_2)\phi^{(H)}(p_3)+
{\cal G}(p_3;p_1,p_2;\omega)\,\phi^{(H)}(p_3)a_{\mu_1}(p_1)a_{\mu_2}(p_2)\big]+}\\[4pt]
{\;\;\;\;\omega^{\mu\nu}\omega^{\rho\sigma}\,(p_2+p_3)_{\nu}\,{p_3}_\sigma\delta^{\mu_1}_\mu\delta^{\mu_2}_\rho\times}\\[4pt]
{\;\;\big[{\cal G}(p_1;p_2,p_3;\omega)\,a_{\mu_1}(p_1)a_{\mu_2}(p_2)\phi^{(H)}(p_3)+
{\;\;\cal G}(-p_1;p_2,p_3;\omega)\,a_{\mu_2}(p_2)\phi^{(H)}(p_3)a_{\mu_1}(p_1)+}\\[4pt]
{\;\;\;\overline{{\cal G}}(p_1;p_2,p_3;\omega)\,a_{\mu_1}(p_1)a_{\mu_2}(p_2)\phi^{(H)}(p_3)+
\overline{{\cal G}}(-p_1;p_2,p_3;\omega)\,a_{\mu_1}(p_1)\phi^{(H)}(p_3)a_{\mu_2}(p_2)\big]-}\\[4pt]
{\;\;\;\;\frac{1}{2}\,\omega^{\mu_1\mu_2}\times}\\[4pt]
{\;\;\big[{\cal F}(p_1;p_2,p_3;\omega)\,a_{\mu_1}(p_1)a_{\mu_2}(p_2)\phi^{(H)}(p_3)+
{\;\;\cal F}(-p_1;p_2,p_3;\omega)\,a_{\mu_2}(p_2)\phi^{(H)}(p_3)a_{\mu_1}(p_1)\big)+}\\[4pt]
{\;\;\;\overline{{\cal F}}(p_1;p_2,p_3;\omega)\,\phi^{(H)}(p_3)a_{\mu_2}(p_2)a_{\mu_1}(p_1)+
\overline{{\cal F}}(-p_1;p_2,p_3;\omega)\,a_{\mu_1}(p_1)\phi^{(H)}(p_3)a_{\mu_2}(p_2)\big]\Big\}.}
\end{array}
\end{equation*}
In the previous equation, $\overline{{\cal G}}$ and $\overline{{\cal F}}$ are the complex conjugates of the functions ${\cal G}$ and ${\cal F}$, respectively. The functions ${\cal G}$ and ${\cal F}$ are defined as follows
\begin{equation*}
\begin{array}{l}
{{\cal G}(p_1;p_2,p_3;\omega)=\frac{1}{p_2\wedge p_3}\;\Big[\frac{e^{-\frac{i}{2}(p_1\wedge p_2+p_1\wedge p_3+p_2\wedge p_3)}-1}{p_1\wedge p_2+p_1\wedge p_3+p_2\wedge p_3}-
\frac{e^{-\frac{i}{2}(p_1\wedge p_2+p_1\wedge p_3)}-1}{p_1\wedge p_2+p_1\wedge p_3}\Big],}\\[4pt]
{{\cal F}(p_1;p_2,p_3;\omega)=\frac{e^{-\frac{i}{2}(p_1\wedge p_2+p_1\wedge p_3+p_2\wedge p_3)}-1}{p_1\wedge p_2+p_1\wedge p_3+p_2\wedge p_3}.}
\end{array}
\end{equation*}

\section{Acknowledgements}
This work has been financially supported in part by MICINN through grant
FPA2011-24568.

\end{document}